\documentclass[12pt,a4paper]{article}
\pdfoutput=1
\newcommand{\be}{\begin{equation}}
\newcommand{\ee}{\end{equation}}
\newcommand{\bea}{\begin{eqnarray}}
\newcommand{\eea}{\end{eqnarray}}
\newcommand{\beas}{\begin{eqnarray*}}
\newcommand{\eeas}{\end{eqnarray*}}
\newcommand{\ba}{\begin{array}}
\newcommand{\ea}{\end{array}}
\newcommand{\nn}{\nonumber}

\newcommand{\bt}{\begin{table}}

\newcommand{\vsi}{\varsigma}

\newcommand{\al}{\alpha}
\newcommand{\ga}{\gamma}

\newcommand{\Ga}{\Gamma}	

\newcommand{\de}{\delta}
\newcommand{\De}{\Delta}

\newcommand{\ka}{\kappa}
\newcommand{\la}{\lambda}
\newcommand{\La}{\Lambda}
\newcommand{\na}{\nabla}

\newcommand{\si}{\sigma}

\newcommand{\g}{\sqrt{-g}}

\begin{document}
\title{
\vspace{-5ex} 
{\bf 
Minimal  quartet-metric gravity: beyond LCDM 
per  scalar graviton as  the unified\\   dark matter  and  dark  energy 
}}
\author{Yury~F.~Pirogov
\\
\small
{\em 
SRC  Institute for High Energy Physics of
NRC Kurchatov Institute,
 Protvino 142281, Russia 
}
}
\date{}
\maketitle
\begin{abstract}
\noindent
In the framework of the minimal quartet-metric gravity/{\em
systogravity},  a~scalar graviton/{\em systolon}  is stated   as a  universal 
dark component, with supplementary manifestations  in the different contexts
either as  dark matter or  dark energy.  An ensuing extension to the  standard
$\Lambda$CDM model is developed.  A modification of the late expansion of  the
Universe,  with an   attractor  of a scalar master equation defining  an
effective cosmological constant,  which supersedes  the true one, is proposed. 
A  new partial solution to  the cosmological constant problem  is discussed. 
\end{abstract}

\section{Introduction }

The present-day standard scenario for evolution  of the Universe is given   by
the $\La$CDM model  or, otherwise,  the standard cosmological  model (SCM).
The scenario  includes  a~prime\-val inflation  followed by the 
 thermodynamic  expansion  with the subsequent  late acceleration 
due to a cosmological constant (CC). 
Despite the impressive  successes of SCM, the evergrowing
evidences  from the astrophysical  observations
may indicate  a vital necessity of going beyond such the
model  in the future.\footnote{For the primeval inflation, see,
e.g,~\cite{Much}. For  the standard $\La$CDM model and beyond, see,
e.g,~\cite{BSC}.} 
SCM  is based on General Relativity  (GR) as a working tool. Thus, going beyond
SCM may, in particular, imply going beyond GR.\footnote{For the gravity beyond
GR, see, e.g.,~\cite{BGR}.} 
In this vein, in Ref.~\cite{Pir1}  there was developed 
an  extension to GR, the so-called   {\em quartet-metric} GR (QMGR),
describing in a general case 
the scalar-vector-tensor gravity in an explicitly  generally covariant (GC)
fashion.  This  theory was proposed for
unifying the tensor gravity with  a  gravitational  dark content (DC), 
the latter manifesting itself  as  the  dark matter (DM) and
dark energy (DE). In the full theory,   DC consists of  the three dark 
components: the   (massive) scalar, vector  and tensor gravitons. It is
reasonable however to start investigating DC  restricting himself  just by  the 
massive scalar graviton  in addition to the conventional  (massless) tensor
one. In such the minimal   version, 
the theory  was 
previously 
applied in the context of  DM for some
cosmic dark structures (DSs)~\cite{Pir2}. In
the present paper, the  minimal  version of the theory  is further elaborated  
in the context of DE for  the Universe as a whole.

In Sect.~2,  the minimal  QMGR, treated as a part  of  the full theory,
is concisely  presented, with the aim of  applying it in
cosmology as a basic gravity theory superseding GR.  The appearance in  the
effective scalar-graviton potential, in excess of the  Lagrangian  potential, 
of a spontaneous contribution, which proves to be crucial,   is explicitly
demonstrated. 
In Sect.~3, the  scalar graviton  is shortly recapitulated   as DM and 
worked-out in more detail as DE.  An extension to the
standard $\La$CDM model   due to  the  omnipresent in the
Universe scalar-graviton field  is
developed. In particular, a  modification of the late
expansion of the Universe, governed  by  an autonomous  scalar 
equation, is proposed. 
It is argued that the general solution   to the equation asymptotically
approaches  an attractor defining, in turn,   an effective
CC  superseding  the true one.  At last, an ensuing hereof   
partial solution to the CC problem is discussed.

\section{Systogravity and scalar graviton}

\subsection{Systogravity Lagrangian}

We start with a concise exposition of the theory of the massive scalar and
massless tensor gravitons  to be  applied in what follows to  the Universe
instead of GR.
As a basic  extension to GR, we  adopt  the effective field theory 
(EFT)  of gravity, QMGR~\cite{Pir1},   given generically by  a GC action:
\be
S = \int L_G(g_{\mu\nu}, X^a, \eta_{ab})\g\, d^4 x .
\ee
Here, an extended gravity    Lagrangian  $L_G$ depends  on the dynamical 
metric field $g_{\mu\nu}$ 
($g\equiv \det (g_{\mu\nu})$) and  a scalar-quartet field $X^a$,
where  $a,b,\dots=0,\dots,3$ are the  indices of the global   Lorentz
symmetry  $SO(1,3)$,  with the  invariant    Minkowski symbol 
$\eta_{ab}$. The latter, in particular, predetermines for consistency the
signature of $g_{\mu\nu}$. The GC scalar fields   $X^a$ are defined up to the
global Lorentz
transformations and shifts $X^a\to X^a+C^a$, with the arbitrary constants $C^a$.
The (piece-wise)  invertible  transformation  in space-time $ \hat x^{\al}
=\de^\al_a X^a(x)$   ($x^\mu=x^\mu(\hat x)$)    presents  the so-called 
(``hidden'') {\em affine} coordinates.\footnote{The special notation
$\al,\beta,\dots=0,\dots, 3$ for
these coordinates is  used just to explicitly distinguish them among all other
ones.} The latter ones are distinguished by the fact that in these coordinates 
an auxiliary affine connection
$\ga^\la_{\mu\nu}(x)$, entering $L_G$ to ensure GC, gets ``hidden'',
$\ga^{\ga}_{\al \beta }(\hat x)=0$ (though, generally,  the
Christoffel connection $\Ga^\ga_{\al\beta}(\hat  x)\neq 0$).\footnote{One
should  not mix the (piece-wise) affine coordinates in a
region  with the locally inertial/Lorentzian  coordinates in an infinitesimal 
vicinity of a point.} 
The affine coordinates are proposed to be  treated physically as comoving 
with the  vacuum.

The most general QMGR Lagrangian of the second  order in the 
derivatives  of metric is presented in~\cite{Pir1}.   Generically, such a
Lagrangian  describes the  scalar,  vector and tensor  gravitons. Imposing some
``natural''  (in a technical sense) restrictions on the Lagrangian parameters,
one can exclude   the  vector graviton as the
most  ``suspicious''  theoretically and phenomenologically,   leaving in  the
leading approximation just  the  scalar graviton   as the  most
``auspicious'', in the line with   the  conventional
tensor gravity.  In what follows, we  
restrict ourselves by such  a  minimal  version of the 
theory.\footnote{Nevertheless, the rest  of the
gravity DC  may prove to be needed 
in the future~\cite{Pir1}.}
Containing   in addition to the 
tensor graviton only  the scalar one or, otherwise, the  {\em
systolon}~\cite{Pir2},  such a
minimal QMGR   from its  particle-content point of view  may for short  be
referred to as  the {\em systogravity}.

The  full  QMGR Lagrangian  $L_G$ reduces  for systogravity  to
\be\label{sg}
L_{sg}\equiv L_g+L_s=-\frac{1}{2}\ka_g^2  R +\frac{1}{2} \ka_s ^2 g^{\ka\la}
\partial_{\ka}\si 
\partial_{\la}\si -V_s(\si).
\ee
Here $R$ is the Ricci scalar, and $\si$  a (dimensionless)
scalar-graviton/systolon field, with $V(\si)$
its potential. 
In the above, $\ka_g$ is the tensor-gravity energy  scale  
given by the reduced Planck mass, $\ka_g=1/\sqrt{8\pi G_N}$, with $\ka_s$ being
a  similar  scale appropriate to the scalar gravity. The respective
physical (dimensionfull)  field is $\vsi\equiv \ka_s\si$. 
For the dominance  of the tensor gravity it is assumed moreover that $
\ka_s/\ka_g\ll1$.
Considering such an EFT at energies less then $\ka_s$, 
we retain  only the leading term in the derivatives of $\si$.  On the other
hand, the scalar potential  $V_s$   is  a~priori an arbitrary function of $\si$.
Nevertheless, the $\si$-dependent part of $V_s$ will be assumed to be
suppressed due to the  approximate shift symmetry $\si\to \si+\si_0$, with any
constant $\si_0$.\footnote{At that, the  smallness of the constant part of the
Lagrangian potential, $V_s|_{\rm min}$, in particular  putting   $V_s|_{\rm
min}=0$, 
remains to be justified as  an additional condition.} 
More particularly,  one has
\be\label{si}
\si= \ln \sqrt{-g}/\sqrt{-\ga},
\ee
where $\sqrt{-\ga}$ is  an auxiliary affine  measure in addition to the
metric one,  $\sqrt{-g}$,  as  follows:
\be\label{key}
\sqrt{-\ga}\equiv X=\det (\partial_\la  X^a) >0,
\ee
  clearly independent of  metric.\footnote{In the
minimal QMGR,  the dependence on $\eta_{ab}$ disappears except  for
defining the signature of $g_{\mu\nu}$.}  With $X^a$
being  the  GC scalars,  $\sqrt{-\ga}$  has the same weight  as 
$\sqrt{-g}$ under the GC transformations.  This   ensures 
$\si$ to be the true scalar field.\footnote{In the
affine coordinates $\hat x^{\al}$, one clearly gets
$\sqrt{-\ga(\hat x)}=1$. Under $\si\neq 0$,  each of the measures,
$\sqrt{-\ga}$ and $\sqrt{- g}$, can always  be brought to unity separately, but
not simultaneously, with a mismatch defining ultimately the scalar $\si$.}
Eqs.~(\ref{si}) and  (\ref{key}) are the  key points of the dynamical  theory
of scalar graviton/systolon.  Had $\si$ been a conventional   elementary field,
this would result in a quite different theory.  On the other hand,  had
$\sqrt{-\ga}$ been a non-dynamical quantity, the theory
would be just the semi-dynamical one~\cite{Pir2},  preserving nevertheless many
features of the fully dynamical theory.\footnote{Due to dependence of
$\ga$ on the derivatives of $X^a$, choosing $\si$  in
(\ref{sg}) as an independent Lagrangian variable, as it might be  tempting
superficially,  is an illegitimate (not a point-wise)  operation  resulting in 
the quite different classical equations. For this reason,
$\si$ can not, generally,  be reduced to an ordinary scalar field.}
For completeness, the pure systogravity Lagrangian  $L_{sg}$ is to be
supplemented by a 
matter one, $L_m$, with  some   matter fields $\Phi^I$.\footnote{In the
hot Universe,  the theory  may, generally,   
describe both  the coherent scalar-graviton/systolon field and the respective 
(incoherent) thermodynamic fraction  with a temperature~$T_s$. We omit  the
latter  fraction (if any) by putting $T_s=0$.}

\subsection{Systogravity field equations}

The  Lagrangian $L_{sg}(g_{\mu\nu}, X)$   is   a marginal
case of a full Lagrangian $L_G(g_{\mu\nu}, X^a)$. For this reason, though  
dependent explicitly only on~$X$,   the Lagrangian 
$L_{sg}$ is understood  as a function of  the whole $X^a$, like the full
$L_G$.\footnote{Beyond the context of $L_G$, the  dependence of $L_{sg}$ on
$X^a$ would be just  a recipe for  choosing the proper  field variables.} The
resulting ambiguity in $X^a$ (at the same $X$)   is inessential
for  systogravity and   may be reduced afterwords  in the full theory.  
By this token, varying the action through  $g_{\mu\nu}$, $X^a$
and $\Phi^I$ one  gets  the system of 
the  coupled  field equations  (FEs)  for the metric, quartet and
matter in the conventional notation, respectively, as  follows:
\bea\label{FEs}
G_{\mu\nu} &\equiv &R_{\mu\nu} -\frac{1}{2}R g_{\mu\nu}=\frac{1}{\ka_g^2}
(T_{s\mu\nu}  + T_{m\mu\nu} ) ,\nn\\
   \frac{\de }{\de X^a}  (L_{s} +L_m)  & \equiv& -  \na_\la\Big((\frac{\de
L_s}{\de \si}
 +  \frac{\de L_m}{\de \si}  )
X^\la_{a}\Big)=0,\nn\\
  \frac{\de  L_m}{\de\Phi^I}  &\equiv& \frac{\partial L_m}{\partial \Phi^I}
-\na^\ka\frac{ \partial L_m}{\partial
\na_\ka\Phi^I}=0,
\eea
where  $X^\la_{a}$ is   a tetrad  inverse to  that $X_\la^a\equiv
\partial_\la X^a$.
The first and the last FEs  in (\ref{FEs}) are clearly the counterparts of
the gravity and matter FEs in GR, while 
the second FE may equally  well be attributed either to the scalar   matter
or to gravity.\footnote{Clearly, this FE encompasses  as a marginal 
case the ordinary solutions for $\si$, with  $\de(L_m+L_s)/\de \si=0$, as
well.}  In the spirit of DC,
the Lagrangian $L_m$ for the ordinary matter  may moreover  be assumed to be
explicitly independent of~$\si$.
In FEs (\ref{FEs}), $T_{s\mu\nu}$ and $T_{m\mu\nu}$ are  the canonical 
energy-momentum tensors, respectively, for the scalar graviton/systolon  and
matter obtained by means of varying  the Lagrangian $L_f$ of a   fraction 
$f=(s,m)$
through $g_{\mu\nu}$ as follows:
\be
T_{f\mu\nu} \equiv   \frac{2}{\g}\frac{\de(\g L_f)}{\de
g^{\mu\nu}}.
\ee
By this token,  one gets
\bea
T_{s\mu\nu}  &= &  \ka_s^2\na_\mu \si \na_\nu \si - \Big(\frac {1}{2} \ka_s^2
\na^\la\si \na_\la\si - U_s\Big) g_{\mu\nu},\nn\\
T_{m\mu\nu}&=&
2\frac{\partial L_m}{\partial g^{\mu\nu}}-\Big( L_m +\frac{\de
L_m}{\de \si}\Big) g_{\mu\nu},
\eea
where  $U_s\equiv V_s+W_s$, with
\be\label{W}
W_s\equiv- \de L_s/\de \si= 
\ka_s^2  \na^\la\na_\la \si +\partial V_s/\partial \si
\ee
being nothing but  the scalar  wave operator.
Formally, the  term $W_s$  appears   in  $T_{s\mu\nu}$ in addition to the
conventional $V_s$ because of   $\si$ being  not independent of~$g_{\mu\nu}$.
Physically,  this off-shell contribution may  be attributed 
to the ``inertia'' of the vacuum revealing ultimately  through the dependence on
$X^a$.\footnote{Because the major part of the energy content of the
Universe is attributed in the $\La$CDM model to DM and DE, associating the
latter ones with the vacuum $X^a$  may realize in
QMGR (with the potential term  also dependent on $X^a$)   a DC counterpart   of
the Mach's principle (irrespective of  the ordinary matter).} 
While  due to GC all  the coordinates, including the affine ones, are
geometrically  equivalent, the physical interpretation in them may look
quite differently   (for the expanding Universe, see, later on).
At last, precisely the off-shell  term $W_s$ crucially distinguishes the scalar
graviton/systolon from its ordinary scalar counterpart,  providing an   ultimate
reason for the ensuing  drastic consequences.\footnote{Such an unconventional 
term   appears in systogravity
like a ``Black Swan''. The latter  may be said, after N.~Taleb,  to be  a
clue event/phenomenon producing drastic consequences  hardly (if any) 
foreseeable  beforehand, but,  in principle,  explainable  afterwards.}

\subsection{Scalar-graviton  effective  potential}

The reduced Bianchi identity, $\na_\nu
G^\nu_\mu=0$, results in the continuity condition, $\na_\nu
T^\nu_\mu=0$, for the total
energy-momentum tensor $T_{\mu\nu}= T_{s\mu\nu}+T_{m\mu\nu}$,  to  be
presented as   follows:
\be\label{Bianchi}
\partial_\mu W_s +W_s \partial_\mu \si 
+\na_\nu T_m{}^\nu_{\mu}=0.
\ee
With account for  (\ref{W}) the equation above  is nothing but the  third-order
FE for $\si$,  being quite complicated.
However, it greatly simplifies if  $T_{m\mu\nu}$ 
fulfills the continuity condition by itself, $ \na_\nu T_m{}^\nu_{\mu}=0$.
A priori, this would require some fine tuning.  Still, if $L_m$ is
independent of $\si$ (and thus of $X$),  this 
condition is  ensured  without  such a  tuning  as in GR  due to   GC (which 
now accounts only  for the  metric and matter, but not for $X$) 
supplemented by the matter FE.
In this case (or under  the  absence of matter)  one gets the   solution to
(\ref{Bianchi}) as follows:
\be\label{W_s}
W_s=W_0 e^{-\si}, 
\ee
with $W_0$ an integration constant.
With account for (\ref{W}) the scalar  FE now becomes
\be\label{W0}
\ka_s^2  \na^\la\na_\la \si +\partial \bar V_s/\partial \si=0,
\ee
where   $\bar V_s$ is   the   effective scalar potential: 
\be
\bar V_s\equiv V_s+W_0 e^{-\si}.
\ee 
In distinction with the parameters of $V_s$,  the
value and  sign   of $W_0$ is not fixed ones forever, but 
is free to vary for the different  solutions.  Eventually,
this arbitrariness allows one to consider  the  cosmic phenomena on the quite
different distance scales starting
from the scales of the scalar-modified black holes (BHs) up to the Universe as a
whole (see, below).

At last,   if $L_m$ is independent of $\si$ (and thus of $X$),
one can,   with account for (\ref{si}) and (\ref{W_s}),   present  FE
(\ref{FEs}) for $X^a$   at $W_0\neq 0$ in the form
\be\label{Qa}
 \partial_\la( \g e^{-\si}X^\la_{a})  =   \partial_\la( \sqrt{-\ga}
X^\la_{a})=0,
\ee
which proves  to be independent of $g_{\mu\nu}$. Thus, after
finding  from FEs the metric and $\si$,  and extracting
hereof  $\sqrt{-\ga}$, one can  find
the proper $X^a$  up to a residual freedom
consistent with  the required $\ga$.
Such an ambiguity  is insignificant in systogravity  and  can, in
principle, be reduced  afterwards in the full theory.  The proposed  minimal
QMGR/systogravity  is   a theoretically  consistent next-to-GR theory of gravity
and, as such,  may be tried  phenomenologically to supersede GR, in particular, 
in cosmology.

\section{Beyond $\La$CDM  per  scalar graviton}

\subsection{Scalar graviton  as  dark component}

Let   the total scalar-graviton/systolon field  $\si(t, {\bf x})$  in the
cosmic coordinates $x^\mu=(t, {\bf x})$ (see, below)  in the Universe be
partiteted as
\be
\si=\si_0(t) + \sum \De\si (t,{\bf x}).
\ee
Here $\si_0$ is an   omnipresent  in the whole Universe (background) 
field and $\De\si$
are the  piece-wise  (perturbation) fields corresponding to a cosmic DSs 
(such as the scalar-modified BHs, galaxies or  the cluster of galaxies)  
appearing on the distance scales much less than that of the Universe.
In the framework of systogravity,  DE  and DM are proposed to be
treated as the supplementary
manifestations of  the same generic field $\si$ as the universal DC in the 
different contexts, respectively, either as    the spatially homogeneous 
$\si_0(t)$ or as  the spatially inhomogeneous $\De\si(t, {\bf x})$.\footnote{A
principle difference between the two
parts  of $\si$  is that $\partial_\mu\si_0$ is taken to be time-like, whereas 
$\partial_\mu\De\si$ is  supposed (at least presently) to be  space-like (as for
the static $\De\si({\bf x})$).} Being a part of the gravity field   in
line with the metric,  $\De\si$ should be treated together  with  the latter  
in the process of the growth of perturbations in the
Universe. To get an idea of a more detailed  picture of $\si$ as   DC,  we
consider  here  only  some simplified cases.

\subsection{Scalar graviton as dark matter}

In the context of DM, we  mention  the quasi-stationary cosmic DSs  in the
present time on the time intervals much less then the Universe evolution time,
neglecting by their  temporal dependence.
The simplest such DSs correspond to 
the static spherically symmetric fields $\De\si(r)$,  where
$r$ is the radial distance  from the  spatial
origin inside the  respective DS, and 
$W_s=W_0 e^{-\De\si(r)}$, $W_0\leq 0$. One may envisage the  three
particular cases of such the DSs (in neglect by the Lagrangian  potential
$V_s$)\footnote{The latter is assumed here  to be   negligible becoming 
important only to  the cut-off of DSs on their
periphery. An opposite case with  the heavy scalar gravitons/systolons
would require a  special consideration.}
as follows~\cite{Pir2}.\footnote{The rotating cosmic DSs,
with the stationary axisymmetric metric 
$g_{\mu\nu}(r,\theta)$, $\theta$ the azimuthal angle,  and the static scalar 
field  $\De\si(r,\theta)$, could a priori be be envisaged, too.}$^,$\footnote{To
be more precise, the considered static solutions refer to the properly chosen
spatial coordinate to be adjusted eventually to the cosmic ones.}

\paragraph{({\em i})} 

The simplest  case  is presented by  the so-called {\em dark  fractures}
(DFs), which are the singular in the
spatial origin (but regular  at the spatial infinity)  compact DSs
corresponding to $W_0=0$.\footnote{Under the latter condition, $\De\si$  is
similar to an ordinary scalar field (but for  the nature of the singularity
in the spatial origin).} 
Being filled up exclusively by the
scalar gravitons/systolons as DM,  DFs are   produced ultimately due to a 
singularity of the space itself in the spatial  origin of the  DFs. In a more
general case, DFs are to be supplemented by the  ordinary matter in their
spatial origin.\footnote{At that, a  singularity in  space may serve as a seed
for the   DF.}
Asymptotically,  DFs mimic DHs  of~GR, with
the total mass determined both by the matter and the scalar field.
In the  limit  $\De\si=0$,  the DFs reduce  to the ordinary BHs.

\paragraph{({\em ii})}   
  
An opposite case  is presented by the so-called {\em dark halos} (DHs),  which
are the  regular in the spatial origin (but singular at the spatial infinity) 
extended DSs. They  correspond  to $W_0< 0$, with the negative
spontaneous term producing an attractive  potential well.
Designating $W_0\equiv-  \ka_s^2/R_0^2$, one gets  $\De\si \simeq
(r/R_0)^2$ at $r<R_0$
and $\De\si\sim  \ln  (r/R_0)^2$ at $r> R_0$,  with $R_0$  the 
soft-core scale.\footnote{This logarithmic  growth should ultimately  be cut-off
by the
scalar-graviton/systolon  mass $m_s$,  with $\la_s=1/m_s\gg R_0$.}  The
equivalent DM 
energy  density for  such the  static spherically symmetric DHs, 
${\rho}_{DH}={\rho}_s+3 p_s $,  
at the distances $r>R_0$   proves to be  ${\rho}_{DH} \sim  \ka_s^2 /
r^2>0$, with the equivalent mass inside a  large  radius 
$R>R_0$ being $M\sim \ka_s^2 R$.
At that, the attractive gravity force  acting at  the distance $R$ on a
test body possessing  a transverse rotation velocity $v$ satisfies  the
relation $(\ka_s/\ka_g)^2/ R\sim v^2/R$.
The DHs are, in principle,  apt  to describe the anomalous rotation curves
with the constant asymptotic velocity  $v|_{\infty}\sim 
\upsilon_s\equiv \ka_s/\ka_g$.\footnote{To get the realistic rotation profiles
in the galaxies, the ordinary matter should also be accounted for. 
Nevertheless, under  $v|_{\infty}\sim
10^{-3}$ and $\ka_g\simeq 2.4\times 10^{18}\,{\rm GeV}$, it is expected that 
$\ka_s\sim 10^{15}\,{\rm GeV}$, which   remarkably proves  to be of the order of
the GUT scale~\cite{Pir2}.}

\paragraph{({\em iii})} 

An interpolating  case is presented by the so-called {\em dark lacunas} (DLs),  
which are the compact-extended DSs corresponding to $W_0< 0$, 
with DFs in the spatial origin   supplemented by DHs at the spatial periphery.
Being the typical DSs, the DLs reduce as the marginal cases  either to DFs or to
DHs  depending on the relation between  the central
singularity and the soft core.   By their spatial configuration,  such the
cosmic DSs may naturally serve as the
``seeds'' for the galaxies.
Supplemented, in turn,  by the ordinary matter,  DLs may be 
expected to constitute the real galaxies on   the various  distance
scales~$R_0$.  At the  much larger  $R_0$,  the DLs
properly modified by matter should, in principle, have the similar  bearing  to
the  clusters of galaxies.\footnote{Some difference may
be expected due to  cut-offs.} To achieve this, the freedom of
arbitrary choosing the spontaneous terms $W_0\sim - 1/R_0^2$ is 
crucial. An ultimate ability (if any) to describe   the  realistic  galaxies
and the galaxy clusters by means of DLs with  matter (and, conceivably,  
rotation) would be crucial for the minimal QMGR/systogravity.

\subsection{Scalar graviton as  dark energy}

\paragraph{
Universe  evolution equations}

The FRW metric  for the homogeneous isotropic Universe  is given in the
conventional notation by the line element
\be
d s^2= d t^2- a^2\Big (\frac{1}{1-K r^2} d r^2 +r^2 d\Omega^2\Big),
\ee
where $t$ is the cosmic standard  time, $r$  the  radial distance from an 
(arbitrary chosen) spatial origin,
$a(t)$ a scale factor, $K=k/l_0^2$, with $l_0$ an arbitrary fixed  unit of
length,  and $k=0,\pm 1$ for the spatially flat, closed and open Universe,
respectively.  Let  the  homogeneous and isotropic Universe  be filled  up by a
continuous medium with  the energy-momentum tensor 
\be
T^{\mu\nu} =(\rho+p) u^\mu u^\nu - p g^{\mu\nu},
\ee
where   ${\rho}$ and $p$ are,  respectively,  the energy density
and pressure,  and $u^\mu$ ($u^\la u_\la =1$)  the comoving four-velocity, with
$u^\mu=(1,0,0,0)$ in the cosmic standard coordinates. Then 
the Friedman-Lema\^itre  equations   for the evolution of the Universe   look
like
\bea\label{Fri}
\ddot a/a &=& - ({\rho}+3p) / 6\ka_g^2  ,\nn\\
H^2+K/a^2&=&   {\rho}/3 \ka_g^2,
\eea
with $H\equiv \dot a/a$   being the Hubble parameter, and  a dot meaning a time
derivative.

To proceed further, we  adopt first of all that the whole Universe 
is  filled up   by the homogeneous  non-stationary  scalar-graviton/systolon  
field $\si_0(t)$.   At the energy scales greater then
$\ka_s$ (but still  lower then $\ka_g$),  the   systogravity  in the
leading order in   $\ka_s^{-1}\partial/ \partial t  $ should be superseded by
EFT  accounting for all orders   of  the latter term (as well as 
dependent, possibly, on the temperature  $T_s$), but still  in the leading
order in $\ka_g^{-1}\partial/\partial t$.  In the spirit of  SCM,
this stage should  describe the primeval inflation.  We have  little to add to
this point,   but to accept it  for granted. Nevertheless,   after inflation, 
at the  energies   less than~$\ka_s$,  the leading-order
systogravity Lagrangian~(\ref{sg}) should  be (approximately)  applicable.
Associating   the homogeneous scalar   field  
$\si_0(t)$ with DE  one can get (omitting here and  in what
follows the  subscript  for~$\si_0$):\footnote{The  role of $u^\mu$ here 
plays $n^\mu= \na^\mu \si/( \na^\la \si  \na_\la \si)^{1/2}$, which in the
cosmic standard coordinates looks like $n^\mu =(1,0,0,0)$. In fact, this may be
taken  in systogravity as a dynamical definition of $u^\mu$.} 
\be\label{DE}
{\rho}_{DE}( p_{DE} )=\frac{1}{2}\ka_s^2  \dot \si^2  \pm U_s ,
\ee
In the  above,  one has $U_s=V_s+W_s$, with the scalar  wave operator 
$W_s$ as follows:
\be\label{Ws}
W_s=\ka_s^2(\ddot\si+3H  \dot \si)+  \partial V_s/\partial\si .
\ee
These expressions are valid at any  $k$ 
and  correspond to the DE effective equation
of state ${\rho}_{DE}=w_{DE} p_{DE}$, with the variable  index $w_{DE}(\si)$.
Under  $\dot \si=0$ (though, generally, $\ddot \si\neq 0$) one has 
$w_{DE}=-1$
mimicking thus a ($\si$-dependent) effective  $\La$-term. 
The appearance in ${\rho}_{DE}$ and  $p_{DE}$
 of the second time derivative of $\si$  through $W_s$  is quite unconventional
feature peculiar to the scalar graviton/systolon due to dependence of $\si$ on
$g_{\mu\nu}$. Had $\si$ been  an ordinary  scalar
field, corresponding to $W_s=0$,  such an  effect  would be  absent.
For this reason, systogravity can not, generally,  be reduced
to GR plus an ordinary scalar field.
Further, as   in the $\La$CDM model  the ordinary matter $m$ and  DM,
generically $M=(m,DM)$,  concentrated in galaxies and the
galaxy clusters, are  to be   smeared out homogeneously all over the
Universe, too. At that,  their  energy densities and pressures  are assumed
satisfying the proper 
equations of state $p_m=w_m {\rho}_m$ and $p_{DM}=w_{DM} {\rho}_{DM}$,
with  the  state indices  $w_{m}$   and $w_{DM}$, respectively. Due to the
relative transparency of galaxies, $p_{DM}$ may be assumed to be small,
justifying thus the assumption of  the cold or warm DM (at
least at  the relatively late stages of the evolution of the
Universe).\footnote{Under the latter assumption, what concerns the
scalar-graviton/systolon DE  becomes, in essence, independent
of the previously  made statement  that DM has the similar nature,  and may be
applicable in a more general context.}
Altogether, the total energy density and pressure, ${\rho}$ and $p$,  are  given
by the sum of the three generic fractions:
\bea\label{tot}
 {\rho} &=& {\rho}_m+  {\rho}_{DM}+  {\rho}_{DE}\equiv   {\rho}_{M}+ 
{\rho}_{DE} ,\nn\\
 p&= &p_m+  p_{DM} + p_{DE} \equiv p_{M} + p_{DE},
\eea 
with $M=(m, DM)$ referring to  the total matter.\footnote{The  more conventional
types of DM (if any) are  counted here  formally as $m$.}
This should be supplemented by 
the  continuity condition  for DE and  the total matter $M$:
\be\label{D}
  \dot W_s  + W_s\dot \si= -(  \dot {\rho}_{M}+3 H({\rho}_{M}+ p_{M})),
\ee
which   fallows from the reduced Bianchi  identify. 
Generally,  this is the third-order equation for $\si$ accounting, in
particular,  for the transition of  DE into  DM (and v.v.). 

Eqs.~(\ref{Fri})--(\ref{D}) present the general
systogravity scenario for  the evolution of the homogeneous isotropic Universe
filled up with the scalar gravitons/systolons and matter. Having found $\si(t)$
and $a(t)$ (and, thus, $\ga(t)$)  one can  then get   with account for
(\ref{Qa}) the   affine quartet $X^a=(X^{0}, X^{A})$,
$A=1,2,3$,  looking at $k=0$ like
\be
X^{0}=\int  \sqrt{-\ga}d t ,\ \ X^{A}=\de^{A}_n x^n
\ee 
($n=1,2,3$), so that $X= \det(\partial_\la X^a)= \sqrt{-\ga}$, as it should be.
This determines, in particular,  the cosmic  affine  time  $\hat t\equiv
X^{0}(t)$ satisfying   $\sqrt{-\hat \ga}  d \hat t= \sqrt{-\ga}d t$, with
$\hat\ga\equiv \ga(\hat t)=-1$.

\paragraph{Effective cosmological constant}

A significant simplification occurs if the total  matter $M= (m,DM)$
is covariantly conserved, with the r.h.s of
(\ref{D})  being zero.\footnote{This suffices for the following. Assuming the 
absence of interactions of the ordinary  matter and DM, one
may moreover consider  their  separate covariant conservation.}
Then it follows hereof that $W_s=W_0 e^{-\si}$,
with $W_0$ an integration  constant. With account for (\ref{Ws}),   this, in
turn, implies that  the scalar-graviton/systolon  FE   at any $k$  are 
determined by the effective
potential $\bar V_s =V_s+W_0 e^{-\si}$ as follows:
\be\label{U_s}
\ka_s^2(\ddot \si+3 H  \dot \si)+\partial \bar V_s / \partial \si =0.
\ee
Let $\bar\si$ be the position of the minimum of the effective potential,
$\partial \bar V_s/\partial \si|_{\bar\si}=0$.
Neglecting   by $\dot\si$ and  $\ddot\si$ reduces this FE  to $\partial
\bar V_s / \partial \si =0$, meaning  $\si$  to be  restricted by
$\bar\si$. 
By this token,  designating
$\bar V_s|_{\bar\si} \equiv \ka_g^2 \bar \La_s$ and  replacing in
${\rho}_{DE}$ and $p_{DE}$ the $\si$-dependent $\bar V_s$ by the constant
$\bar V_s|_{\bar\si}$, 
one arrives  (assuming  $w_{DM}=0$) at the standard $\La$CDM
model,  corresponding to  the  effective  CC  $\bar \La_s$, so  that 
\be
\bar{\rho}_{DE}=- \bar p_{DE}=\ka_g^2\bar\La_s.
\ee
There are nevertheless  two important caveats. First,  being determined through
an interplay of the given  Lagrangian  potential $V_s$ and the  spontaneous  
contribution $W_0$,   the effective CC is not  a true  fundamental
parameter.\footnote{In particular, one may envisage the situation when a 
counterpart of the Lagrangian CC given by $V_s|_{\rm min}\equiv
\ka_g^2\La_s$,   is zero,  whereas the effective CC, $\bar \La_s$, is finite 
due to $W_0\neq 0$, solving thus  partially  the CC problem. This   resembles 
the  situation  in  Unimodular Relativity (UR), where  CC is not  the
Lagrangian parameter $\La$, but an integration constant~$\La_0$ appearing
spontaneously at the level of   FEs.}  Second,
the effective $\La$CDM  model is valid  for $\si$ only 
in a relatively narrow region around $\bar\si$,
being superseded in a wider region by the systogravity
extension to SCM. 
The deviations from SCM  may be crucial  when $\si$ is far away 
from~$\bar\si$.  A number of uncertainties may however be envisaged,
e.g., an   omitted leakage of DE into DM (or v.v.), as well as the neglected 
temperature dependence of the theory, etc. Nevertheless, reproducing in a limit 
the standard $\La$CDM model,  systogravity hopefully  provides 
a proper way to extend  such the  (conceivably, simplified) model.

\paragraph{Modified late expansion}

Restricting  consideration by  the late stage of the  evolution of
the Universe,  we   assume  at this stage the dominance of DE
by putting ${\rho}_M=p_M=0$. 
With account for  (\ref{Fri}),  Eq.~(\ref{U_s}) gets   at $k=0$ the autonomous 
form as follows:
\be\label{PS}
\ddot \si+\sqrt{3}   \Big( \frac{1}{2} \upsilon_s^2   \dot \si^2
+ \bar V_s/\ka_g^{2}  \Big)^{1/2} \dot \si+\ka_s^{-2}\partial \bar V_s /
\partial \si =0,
\ee
where $\upsilon_s=\ka_s/\ka_g\ll 1$.  This is the master equation for
the  evolution of the  Universe due to the scalar-graviton/systolon  DE. Having
found hereof $\si$ one then finds $H\equiv\dot a/a={\rho}_{s}^{1/2}/
\sqrt{3}\ka_g \ge 0$, as well as   the respective scale factor
\be\label{a}
a=a_0 \exp\frac{1}{\sqrt{3}}   \int \Big(  \frac{1}{2} \upsilon_s^2\dot
\si^2 + \bar V_s/ \ka_g^2 \Big )^{1/2}d t  ,
\ee
with $a_0$ an integration constant. To envisage  the behaviour
of $H$  we note that
from (\ref{Fri}) at $k=0$ one can find 
\be
\dot H = - \frac{1}{2}\upsilon_s^2 \dot \si^2\leq 0, 
\ee independent of $\bar V_s$. This means, in particular, that the Hubble 
parameter $H$  always monotonically decays approaching from above a constant
value~$\bar H\ge 0$, which  however may  depend on~$\bar V_s$.  At that, the
Hubble horizon   $H^{-1}$   monotonically expands to $\bar H^{-1}$.

In distinction with  DM, which is dominated
by the kinetic scalar-graviton/systolon contribution and  weakly depends  on 
the potential, the respective DE may crucially  depend on  $V_s$. To
be more specific, let us choose the quadratic  $V_s$, so that
\be
\bar V_s=\frac{1}{2}Q_s \si^2+W_0 e^{-\si},
\ee
where $Q_s\equiv m_s^2 \ka_s^2$, with $m_s$ the scalar-graviton/systolon
mass.
Considering the dependence of  the minimum of $\bar V_s$  on $W_0$ at a fixed
$Q_s$ one roughly gets  for $\bar \si$ and  $\bar
V_s|_{\bar \si}\equiv \ka_g^2\bar\La_s$:
\bea
\bar \si \simeq W_0/Q_s,  \ \   \ka_g^2\bar\La_s \simeq   W_0 , &&
{\rm at} \ \   W_0/Q_s \ll 1,\nn\\
\bar \si\sim 1,  \  \  \ka_g^2\bar\La_s \sim W_0\sim Q_s, &&  {\rm at}\ \
W_0/Q_s\sim 1,\nn\\
\bar \si \simeq\ln W_0/Q_s, \ \ \ka_g^2\bar\La_s\simeq  \frac{1}{2} Q_s 
\ln^2 W_0/Q_s, && {\rm at } \ \ W_0/Q_s \gg 1.
\eea
Clearly, $\si\equiv\bar \si$ is the exact solution to Eq.~(\ref{PS}).
Moreover, studying the  latter in the phase plane ($\si, \dot\si$) 
shows that  $(\bar\si, 0)$ is an  attractor for the solutions which wind 
around  the attractor approaching the latter asymptotically at  $t\to\infty$.
It  then follows from~(\ref{a})  that at $k=0$  there fulfills 
$\bar a =a_0  \exp \bar H t $.
with   $\bar H= (\bar\La/3)^{1/2}$. 
This indicates the appearance  in systogravity  of 
 the modified late  inflation even under the absence of the  true CC.
Had any of $W_0$ or $Q_s$ been
zero the asymptotic inflation would not take place (under the assumed $V_s|_{\rm
min}=\ka_g^2\La_s=0$).
Having explained/assumed  the  exact vanishing  of the true CC, one can
eventually explain   in the  framework of systogravity the effective CC  to be
not exactly zero, solving thus partially  the
CC problem.\footnote{It may be said that systogravity
realizes   for  the Universe, dominated by  the scalar-graviton/systolon DE,  an
interplay of the predetermination  (through the Lagrangian $V_s$)  and the 
occasion (though the spontaneous $W_0$). In particular, 
this would  distinguish  the late  fates of the members
of an ensemble of the ``multiverses'' (if any).}

For  completeness,  the cosmic 
affine time  for the attractor  at $k=0$ is  $\hat t =\bar
X^{0}=\bar t_0 \exp 3\bar H t$,  $\bar t_0$ an integration  constant,  or
inversely $t= (3\bar H)^{-1} \ln \hat t/\bar t_0$,  so that 
$d s^2=  d  \hat t^2/ (3\bar H  \hat t)^2-
(3e^{\bar\si}\bar H \hat t)^{2/3} d {\bf x}^2$, where   $\bar t_0$
is properly chosen to ensure $e^{\bar \si}= \sqrt{-\bar g}$ with $\sqrt{-\bar
\ga}= 1$. In these terms, the exponential scale factor reduces to the power one,
$\bar a\sim (3\bar H \hat t)^{1/3}$. This implies, in particular, that
the cosmic standard time $t$ and  the  scale factor $\bar a$ during the late
stage of the Universe expansion decelerate 
in terms of $\hat t$. Thus using    the  (physically distinguished) comoving
with the vacuum cosmic affine time   as the   reference one would drastically
change the view on the evolution of the Universe.\footnote{Adopting the
affine time may be in concord with the Fock's insistence (in the  context
of a scalar field in GR) on the importance of choosing the proper  coordinates
in GR, as the  GC  geometrical theory,  for better clarifying its physical
content~\cite{Fock}.}

\section{Conclusion}

It may be  concluded that  in  framework of   the minimal QMGR/systogravity  the
omnipresent scalar-graviton/systo\-lon  field filling  up homogeneously 
the Universe, with  the (quasi-)stationary lacunas of the field  as
the  seeds  for   the galaxies or  the  cluster of galaxies, may,  in
principle, present a concise picture of the Universe. 
Reproducing in a limit  the  standard $\La$CDM model,  the ensuing
in  systogravity  scenario may eventually supersede 
the standard one. In particular, there  fallows hereof an alternative
approach to CC  as the  effective one determined by the interplay of the
Lagrangian and spontaneous contributions to the effective scalar  potential. 
More detailed investigations  of  systogravity are
evidently  needed to support/restrict  the proposed extension
to the SCM/$\La$CDM model and   the  alternative  approach to CC.
It seems that the proposed scenario for the evolution of the
Universe  ought to be eventually either confirmed or rejected.
After all,  studying the  minimal QMGR/systogravity as the next-to-GR
theory of gravity   to reveal/exclude the
scalar graviton/systolon  as the  new  particle,  which  drastically
influences the Universe,  is worthy to pursue. Under  success,  this
would pave the way towards the full QMGR and beyond.

\paragraph{Acknowledgement}
Thanks are due  to I.Y.~Polev for assistance with the computer calculations. 
The  stimulating e-mail discussions  with Gilles Cohen-Tannoudji on the Mach's
principle and the CC problem are gratefully acknowledged.

\end{document}